\documentstyle[12pt]{article}
\include{FEYNMAN}
\include{epsf}
\begin{document}
\begin{flushright}
GUTPA/97/06/1\\
NBI-HE-97-27\\
hep-ph/9706428
\end{flushright}
\vskip .1in
\newcommand{\lapprox}{\raisebox{-0.5ex}{$\
\stackrel{\textstyle<}{\textstyle\sim}\ $}}
\newcommand{\gapprox}{\raisebox{-0.5ex}{$\
\stackrel{\textstyle>}{\textstyle\sim}\ $}}

\begin{center}

{\Large \bf
A Natural Solution to the Neutrino Mixing Problem. }

\vspace{50pt}

{\bf C.D. Froggatt and M. Gibson}

\vspace{6pt}

{ \em Department of Physics and Astronomy\\
 Glasgow University, Glasgow G12 8QQ,
Scotland\\}

\vspace{18pt}

{\bf H.B.  Nielsen}

\vspace{6pt}

{ \em The  Niels  Bohr Institute \\
Blegdamsvej 17, DK-2100 Copenhagen {\O}, Denmark \\}
\end{center}

\section*{ }
\begin{center}
{\large\bf Abstract}
\end{center}
The combined requirements, of (i) a natural solution to the fermion mass
hierarchy problem and (ii) an explanation of both the atmospheric and
solar neutrino problems, lead to an essentially unique picture of
neutrino masses and mixing angles. The electron and muon neutrinos
are quasi-degenerate in mass with maximal mixing, giving
$\nu_e - \nu_{\mu}$ vacuum oscillations.
The overall neutrino mass scale is set by the atmospheric neutrino
requirement $\Delta m^2 \sim 10^{-2}$ eV$^2$, implying a mass
for $\nu_e$ and $\nu_{\mu}$ of order \mbox{1 eV} in models with a natural
mass hierarchy, whilst the tau neutrino is expected to be much lighter than
this and only weakly mixed. We present an explicit example based on the
anti-grand unification model of fermion masses.
\thispagestyle{empty}
\newpage

\section{Introduction}

The observed hierarchy of quark and lepton masses and quark
mixing angles strongly suggests the existence of an
approximately conserved chiral
flavour symmetry \cite{fn1} beyond the Standard Model (SM).
For theories in which this chiral symmetry group forms
part of the extended gauge group, the values of the chiral
flavour charges are strongly constrained by anomaly cancellation
conditions. Several models of this type have been constructed
\cite{leurer,ibanezross,binetruyramond,pokorski,fln:np1,SMG3U1}
which give a realistic quark and charged lepton mass
spectrum, without any fine-tuning. In this letter we consider
the structure of the neutrino mass matrix in such models with
a natural mass hierarchy. We show that consistency with
atmospheric and solar
neutrino data can then only be obtained if they are both due to
$\nu_e - \nu_{\mu}$ vacuum oscillations.

As we pointed out some time ago \cite{fn2} the effective three
generation light neutrino mass matrix $M_{\nu}$ in models with
approximately conserved chiral flavour charges, generated for
example by the usual see-saw mechanism, can have two qualitatively
different types of eigenstate. This is a consequence of the
hierarchical structure and symmetry $M_{\nu} = M_{\nu}^T$ of
the mass matrix. In the first case, a neutrino can combine with
its own antiparticle to form a Majorana particle and has small
mixing angles with the other neutrinos. We shall be interested
in the case where the tau neutrino combines with the tau
antineutrino. The second type of
eigenstate corresponds to a neutrino combining with an antineutrino,
which is not the CP conjugate state, to form a 2-component
massive neutrino. Such states naturally occur in pairs with
quasi-degenerate masses and maximal mixing
($\sin^2 2\theta \simeq 1$). We shall be interested in the
case where the electron neutrino combines with the muon
antineutrino; the other member of the quasi-degenerate pair
is formed by combining the muon neutrino with the electron
antineutrino. The fractional mass difference between the two
eigenstates is suppressed by the appproximately conserved
chiral charges ($\Delta m/m \ll 1$).

In the next section, we discuss the structure of the neutrino
mass matrix in models with a natural fermion mass hierarchy.
We then consider the phenomenology of neutrino oscillations in
such models in section \ref{phenomenology}. It is shown that the
only way of obtaining a simultaneous solution of the atmospheric
and solar neutrino problems without fine-tuning is via
$\nu_e - \nu_{\mu}$ vacuum oscillations. An explicit
example based on the anti-grand unification model (AGUT) is
then presented in section \ref{model}. The chiral charges of
the quarks and leptons in the AGUT model are essentially uniquely
determined by the anomaly cancellation conditions. However the
overall neutrino mass scale is not explained in the model, since the
natural see-saw mass scale is set by the Planck mass,
which gives a too low neutrino mass scale of
$\frac{\langle \phi_{WS}\rangle^2}{M_{Planck}}
\sim 3 \times 10^{-6}\ eV$. It is therefore necessary to introduce
the overall neutrino mass scale by hand. This is done by
introducing an effective Higgs field which is a triplet under
the electroweak SU(2) gauge group and assigning it an
ad hoc vacuum expectation value, determined
phenomenologically by the atmospheric neutrino data.

\section{The Fermion Mass Hierarchy}
\label{hierarchy}

The masses of the charged fermions range over five orders of magnitude
from the electron to the top quark. It is only the top quark which has
a mass of order the electroweak scale
$\langle \phi_{WS} \rangle = 174$ GeV
and has a SM Yukawa coupling of order unity. All of the other quark
and lepton masses are suppressed relative to this scale. It is
natural to interpret the different orders of magnitude of the
suppression factors as due to different products of small symmetry
breaking parameters, arising from some approximate chiral gauge symmetry
beyond that of the Standard Model Group (SMG). The SMG is then the
low energy remnant of some larger gauge group G and the SM Yukawa
couplings are effective coupling constants which, in general, are
forbidden by gauge invariance under G. The gauge group G is
supposed to be spontaneously broken to the SMG at some high
energy scale and the effective SM Yukawa couplings are thereby
generated. These suppressed effective couplings of
left-handed to right-handed quarks and leptons are mediated by
vector-like super-heavy intermediate states. If all the
appropriate superheavy states exist, with masses of order the
fundamental mass scale $M_F$ of the extended theory, the
suppression factors are determined by the gauge quantum numbers
of the fermions and the Higgs fields.
In this way charged fermion mass matrices are generated, for
which the different matrix elements can naturally be of
different orders of magnitude and give a realistic mass and
mixing hierarchy.

\begin{figure}[t]
\begin{picture}(39000,13000)
\THICKLINES

\drawline\fermion[\E\REG](3000,7000)[4000]
\global\advance \pmidy by 1000
\put(\pmidx,\pmidy){${\nu}_{\mu}$}

\drawline\fermion[\E\REG](7000,7000)[4000]
\global\advance \pmidy by -1500
\put(\pmidx,\pmidy){$M_F$}

\drawline\fermion[\E\REG](11000,7000)[4000]
\global\advance \pmidy by -1500
\put(\pmidx,\pmidy){$M_F$}

\drawline\fermion[\E\REG](15000,7000)[4000]
\global\advance \pmidy by -1500
\put(\pmidx,\pmidy){$M_F$}

\drawline\fermion[\E\REG](19000,7000)[4000]
\global\advance \pmidy by -1500
\put(\pmidx,\pmidy){$M_F$}

\drawline\fermion[\E\REG](23000,7000)[4000]
\global\advance \pmidy by 1000
\put(\pmidx,\pmidy){$\overline{\nu}_{\tau}$}

\drawline\scalar[\S\REG](7000,7000)[3]
\global\advance \pmidx by 1000
\global\advance \pmidy by -1500
\put(\pmidx,\pmidy){$\phi_{WS}$}
\global\advance \scalarbackx by -530
\global\advance \scalarbacky by -530
\drawline\fermion[\NE\REG](\scalarbackx,\scalarbacky)[1500]
\global\advance \scalarbacky by 1060
\drawline\fermion[\SE\REG](\scalarbackx,\scalarbacky)[1500]

\drawline\scalar[\N\REG](11000,7000)[3]
\global\advance \pmidx by 1000
\global\advance \pmidy by 1000
\put(\pmidx,\pmidy){$T$}
\global\advance \scalarbackx by -530
\global\advance \scalarbacky by -530
\drawline\fermion[\NE\REG](\scalarbackx,\scalarbacky)[1500]
\global\advance \scalarbacky by 1060
\drawline\fermion[\SE\REG](\scalarbackx,\scalarbacky)[1500]

\drawline\scalar[\N\REG](15000,7000)[3]
\global\advance \pmidx by 1000
\global\advance \pmidy by 1000
\put(\pmidx,\pmidy){$W$}
\global\advance \scalarbackx by -530
\global\advance \scalarbacky by -530
\drawline\fermion[\NE\REG](\scalarbackx,\scalarbacky)[1500]
\global\advance \scalarbacky by 1060
\drawline\fermion[\SE\REG](\scalarbackx,\scalarbacky)[1500]

\drawline\scalar[\N\REG](19000,7000)[3]
\global\advance \pmidx by 1000
\global\advance \pmidy by 1000
\put(\pmidx,\pmidy){$T$}
\global\advance \scalarbackx by -530
\global\advance \scalarbacky by -530
\drawline\fermion[\NE\REG](\scalarbackx,\scalarbacky)[1500]
\global\advance \scalarbacky by 1060
\drawline\fermion[\SE\REG](\scalarbackx,\scalarbacky)[1500]

\drawline\scalar[\S\REG](23000,7000)[3]
\global\advance \pmidx by 1000
\global\advance \pmidy by -1500
\put(\pmidx,\pmidy){$\phi_{WS}$}
\global\advance \scalarbackx by -530
\global\advance \scalarbacky by -530
\drawline\fermion[\NE\REG](\scalarbackx,\scalarbacky)[1500]
\global\advance \scalarbacky by 1060
\drawline\fermion[\SE\REG](\scalarbackx,\scalarbacky)[1500]
\end{picture}
\vskip .3cm
\caption{Example Feynman diagram for a neutrino mass
matrix element generated by the see-saw mechanism and
suppressed by the approximately conserved chiral
gauge quantum numbers of the AGUT model.
The crosses indicate the couplings of the Higgs fields to the vacuum.}
\label{numass}
\end{figure}
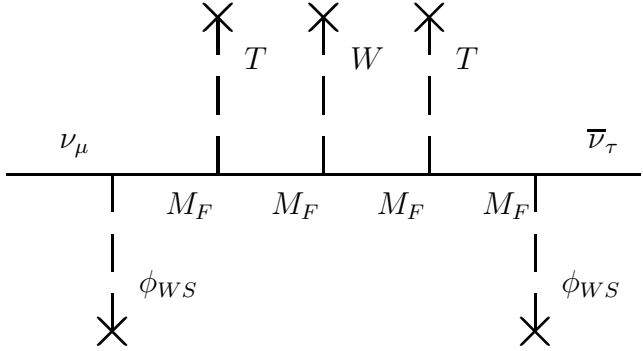

This scheme is readily extended \cite{fn2} to generate a
non-zero light neutrino mass matrix $M_{\nu}$:
\begin{equation}
{\cal L}_{m} = (M_{\nu})_{ij}\nu_{L_i}C\nu_{L_j} +\mbox{h.c.}
\label{Mnu}
\end{equation}
It is then necessary to exchange either the Weinberg-Salam Higgs
field $\phi_{WS}$ twice---the see-saw mechanism
\cite{fn2,seesaw}---as illustrated in fig.~\ref{numass}, or
a weak isotriplet Higgs field $\Delta$ \cite{gelmini}.
As an example, the Higgs field exchanges required to generate the mass
matrix element connecting the left-handed muon neutrino
to the right-handed tau antineutrino, via the see-saw mechanism,
are shown in fig.~\ref{numass} for an AGUT model of the
fermion masses \cite{SMG3U1,fgns}. Assuming all the fundamental
Yukawa couplings are of order unity, this diagram gives the
order of magnitude expression:
\begin{equation}
(M_{\nu})_{\mu\tau} = \frac{\langle \phi_{WS} \rangle^2}{M_F}
\frac{\langle W \rangle}{M_F} \frac{\langle T \rangle^2}{M_F^2}
\label{mutau}
\end{equation}
for the matrix element. The first factor
$\frac{\langle \phi_{WS} \rangle^2}{M_F}$
is the see-saw neutrino mass scale, while
$\frac{\langle W \rangle}{M_F}$ and
$\frac{\langle T \rangle^2}{M_F^2}$ are suppression
factors arising from the exchanges of the Higgs fields,
$W$ and $T$, needed to match the chiral gauge quantum
number differences between $\nu_{\mu}$ and $\overline{\nu}_{\tau}$
in the AGUT model. A similar structure is obtained in other
models with approximately conserved $U(1)$ charges
\cite{dreiner,lavignac}.

As is the case for the charged fermion mass matrices,
the neutrino mass matrix $M_{\nu}$
is determined up to factors of order unity by the quantum
numbers of the neutrinos and Higgs fields, provided we assume
the existence of all the necessary intermediate states at
the fundamental mass scale $M_F$. In some models
this is not true and the superheavy fermion spectrum is
constrained, often by specifying a heavy Majorana
(right-handed neutrino) mass matrix.
The quantum numbers of the SM
neutrino states are of course the same as those of
the charged leptons in the corresponding weak isodoublets.

The neutrino mass matrix $M_{\nu}$ is, by its very definition
eq.~(\ref{Mnu}), symmetric. Also, in models
with approximately conserved chiral $U(1)$ charges,
the matrix elements are generally of different orders
of magnitude due to the presence of various suppression
factors similar to those in eq.(\ref{mutau}). Thus the
generic structure for $M_{\nu}$ is a matrix in which the
various elements typically each have their own order
of magnitude, except in as far as they are forced to be equal
by the symmetry $M_{\nu} = M_{\nu}^T$.
The largest neutrino mass eigenvalue is then given by the
largest matrix element of $M_{\nu}$. If it
happens to be one of a pair of equal off-diagonal elements,
we get two very closely degenerate states as the heaviest
neutrinos and the third neutrino will be much lighter and,
in first approximation, will not mix with the other two.
If the largest element happens to be a diagonal element,
it will mean that the heaviest neutrino is a Majorana neutrino,
the mass of which is given by this matrix element, and it will
not be even order of magnitude-wise degenerate with the other,
lighter neutrinos. These light neutrinos may or may not get
their masses from off-diagonal elements and thus,
in first approximation, be degenerate.

In models with approximately conserved chiral charges,
there is a tendency for a pair of quasi-degenerate neutrinos
to form; these are typically the heaviest neutrinos \cite{fn2}. These
neutrino states may couple dominantly to any pair of charged
leptons. Thus the strongly mixed quasi-degenerate pair are
essentially just as likely to be electron and muon neutrinos
as muon and tau neutrinos or electron and tau neutrinos.

The lepton mixing matrix $U$ is defined analogously to the usual
CKM quark mixing matrix, in terms of the unitary transformations
$U_{\nu}$ and $U_E$, on the left-handed lepton fields, which
diagonalise the squared neutrino mass matrix
$M_{\nu}M_{\nu}^{\dagger}$  and the squared charged lepton mass
matrix $M_EM_E^{\dagger}$ respectively:
\begin{equation}
U = U_{\nu}^{\dagger}U_E
\end{equation}
The charged lepton unitary transformation $U_E$ is expected to
be quasi-diagonal, with small off-diagonal elements due
to the charged lepton mass hierarchy. On the other hand when
there is a quasi-degenerate pair of neutrinos, because off-diagonal
elements dominate their masses, the mixing angle contribution
from $U_{\nu}$ will be very close to $\pi/4$. This is because
then, in first approximation, $U_{\nu}$ has to diagonalise the
$\sigma_x$ Pauli matrix, leading to eigenstates which are 50\%
probability mixtures of two of the original neutrino states.
The lepton mixing matrix $U$ will have a similar structure,
since $U_E$ is quasi-diagonal.
If there is no pair of quasi-degenerate neutrinos, $U_{\nu}$
and $U$ are expected to be quasi-diagonal like $U_E$

So we conclude that there are two generic forms for the neutrino
masses and mixing angles. In the first case there are a pair of
quasi-degenerate neutrinos with essentially maximal mixing and
a third essentially unmixed Majorana neutrino. In the second case
the neutrino spectrum is similar to those of the charged fermion
families, being hierarchical and having small mixing angles.

\section{Neutrino Phenomenology}
\label{phenomenology}

{}From the above discussion we see that models of this type could
generate a neutrino spectrum which has small mixing between all three
neutrinos. However, in order to explain the atmospheric
neutrino problem it is necessary to have large mixing (for
two neutrino mixing $\sin^2(2\theta) \gapprox 0.7$). So we need only
consider the case where we have two nearly degenerate neutrinos
with almost maximal mixing.

There are three possibilities for neutrino mass matrices of this
form; we may have the large mixing between the electron and mu
neutrinos, the electron and tau neutrinos, or the mu and tau
neutrinos. The atmospheric neutrino problem cannot be explained by
$\nu_{\tau}-\nu_e$ mixing so we can immediately discount that scenario.
Hence we are left with the cases of nearly maximal mixing between
$\nu_e$ and $\nu_{\mu}$ or $\nu_{\mu}$ and $\nu_{\tau}$ with the remaining
neutrino mixing only slightly.

In the following we shall refer to
the mass eigenstate neutrinos as $\nu_1, \nu_2$
and $\nu_3$, with corresponding masses $m_1, m_2$ and $m_3$ (defined
as the moduli of the mass eigenvalues). When $\nu_{e}$ and
$\nu_{\mu}$
are strongly mixed these mass eigenstates will be
approximately given in terms of the flavour eigenstates by:
\begin{eqnarray}
\left | \nu_{1} \right \rangle & \simeq &
\frac{1}{\sqrt{2}} \left ( \left | \nu_e \right \rangle
+ \left | \nu_{\mu} \right \rangle \right ) \\
\left | \nu_2 \right \rangle & \simeq &
\frac{1}{\sqrt{2}} \left ( \left | \nu_e \right \rangle
- \left | \nu_{\mu} \right \rangle \right ) \\
\left | \nu_3 \right \rangle & \simeq & \left | \nu_{\tau} \right \rangle
\end{eqnarray}
In the case of large $\nu_{\mu}-\nu_{\tau}$ mixing we obtain
similar relations between flavour and mass eigenstates by making
the replacements $e \leftrightarrow \tau$ and $1
\leftrightarrow 3$ in the above equations.
We also define the mass squared differences by
$\Delta m^2_{ij} = | m^2_i - m^2_j|$.

If we consider the large $\nu_{\mu}-\nu_{\tau}$ mixing scenario
then we must have $\Delta m^2_{23} \sim 10^{-2} \ \mathrm{eV}^2$
for the atmospheric neutrino problem. We also want to explain the
solar neutrino problem, and this requires mixing with the electron
neutrino. The only small mixing solution to this problem is
the MSW solution \cite{MSW} which has $\Delta m^2_{e2(3)} \sim 10^{-5}
\ \mathrm{eV}^2$. Hence we would need:
\begin{eqnarray}
\Delta m^2_{e2(3)} & = & \left| m^2_{\nu_{e}} - m^2_{2(3)} \right|
\sim 10^{-5} \ \mbox{eV}^2 \\
\Delta m^2_{23} & = & \left| m^2_{2} - m^2_{3} \right|
\sim 10^{-2} \ \mbox{eV}^2  \ll m^2_{2}
\end{eqnarray}
where $m_2^2$ is much greater than $\Delta m^2_{23}$
 because we have nearly degenerate
$\nu_{2}$ and $\nu_{3}$.

Clearly the only way we can satisfy these equations is if the
$\nu_e$ is nearly degenerate to  $\nu_2$ and $\nu_3$
(indeed the degree of degeneracy would need to be much greater
to one of them than that between $\nu_2$ and $\nu_3$). This
would require extreme fine tuning since there is no reason to expect the
slightly mixed neutrino to be nearly degenerate with the other
two.

So, we are left with large mixing between $\nu_e$ and $\nu_{\mu}$.
In this case we could solve the atmospheric neutrino problem by
$\nu_e - \nu_{\mu}$ mixing, with $\Delta m^2_{12} \sim
10^{-2} \ \mathrm{eV}^2$, \cite{suzuki}.
 The MSW solution with $\nu_e - \nu_{\tau}$
mixing is again prevented since the fine tuning involved would
again be unnatural. `Just so' vacuum oscillations, where about
one $\nu_e - \nu_{\mu}$ oscillation length lies between the sun and the
earth, require  $\Delta m^2_{12} \sim 10^{-10} \ \mathrm{eV}^2$,
which is clearly incompatible with the atmospheric neutrino
solution. It would seem that we have eliminated all the possible natural
solutions with this type of model.

However, as pointed out in \cite{E-independent},
 recent standard solar model calculations
allow greater freedom in the solutions to the solar neutrino problem.
These calculations vary in their predictions of the $^8B$ flux by more
than a factor of two. If this flux is  treated as a free parameter
within this range then it is possible to get acceptable `energy-independent'
solutions. By `energy-independent' vacuum oscillation solutions we mean that
$\Delta m^2$ is sufficiently large that many oscillation lengths
lie within the sun-earth distance, so that the reduction in
$\nu_e$ flux does not depend on the energy of the solar neutrinos.
Whilst an acceptable solution can be found in this way it should be
noted that changing the $^8B$ flux does not alter the disagreement
between our prediction and the different flux suppressions measured
at KAMIOKANDE and HOMESTAKE \cite{suzuki} and we would still require
these experiments to measure the same flux
suppression factor of $\frac{1}{2}$. Hence
we can now solve the solar neutrino problem with large mixing and
\begin{equation}
10^{-10} \ \mathrm{eV}^2 \lapprox \Delta m^2_{12}
\lapprox 10^{-2} \ \mathrm{eV}^2.
\end{equation}
This is clearly compatible with the atmospheric neutrino solution
if we take $\Delta m^2_{12} \sim 10^{-2} \ \mathrm{eV}^2$. So we
now have an essentially unique solution to the solar and atmospheric
neutrino problems within models of this type.

There is also a controversial indication for neutrino masses from the LSND
experiment, \cite{lsnd}, a $\overline{\nu}_{\mu} \rightarrow \overline{\nu}_e$
appearance experiment. However, a large mixing angle ($\sin^2(2\theta) \sim 1$)
fit
to the LSND data gives
$\Delta m^2_{12} \sim 6 \times 10^{-2} \ \mathrm{eV}^2$ and this is
inconsistent with reactor data \cite{achkar} which for large mixing
require $\Delta m^2_{12} \lapprox 10^{-2} \ \mathrm{eV}^2$. So we would predict
that the LSND result will prove to be unfounded.

It is interesting to note the range of masses we would
expect to arise for $\nu_e$ and $\nu_{\mu}$ in these models.
Since we have a hierarchical structure (apart from the near degeneracies
arising from large off-diagonal elements) it is reasonable to assume
 $\Delta m_{12} \lapprox 0.1 m_1$. We can also use the experimental
limit on the electron neutrino mass $m_{\nu_e} < 15 \ \mathrm{eV}$
\footnote{There are well known problems with the experimental
limit on $m_{\nu_e}$, and it should be noted that this is a fairly
conservative bound; bounds as low as $m_{\nu_e} < 4.4 \ \mathrm{eV}$
 are claimed by some experiments.} \cite{pdg}
to obtain:
\begin{equation}
0.2 \ \mbox{eV} \lapprox m_{1(2)} < 15 \ \mathrm{eV}.
\end{equation}
It also follows from the limit on $m_{\nu_e}$ and
$\Delta m^2_{12} \sim 10^{-2} \mathrm{eV}^2$ that we
would require $\Delta m_{12} = \left | m_1 - m_2 \right |
\gapprox 3 \times 10^{-4} \ \mathrm{eV}$.

{}From the discussion in the previous section we would expect
it to be more usual for the tau neutrino to be lighter than
the other neutrinos. Indeed, since the suppression factors
are due to charge differences which are similar to those in
the charged fermion sector, it seems likely that they will
span a similar range ($\sim 5$ orders of magnitude). If this
is the case then the tau neutrino must be the lightest neutrino
to avoid the cosmological bound for stable neutrinos
$\sum m_{\nu} < 40 \ \mathrm{eV}$.

Since masses with $\sum m_\nu \gapprox 1 \ \mathrm{eV}$ will
significantly contribute to
dark matter we would expect it to be common (but not essential)
for models of this
nature to generate hot dark matter candidates.
The tau neutrino would be much too light to contribute significantly,
so we would expect any dark matter contribution to
come from the electron and mu neutrinos.

So we have an essentially unique solution to the solar and
atmospheric neutrino problems with mass matrices of the
form:
\begin{eqnarray}
M_{\nu}  \sim  \left ( \begin{array}{ccc}
        B_1 & A  &
        X_1 \\
        A & B_2 & X_2 \\
        X_1 & X_2 & C \end{array} \right )
\label{mnu}
\end{eqnarray}
where $A$ is the dominant element and we would have:
\begin{equation}
m_1 \sim m_2 \sim \left | A \right |
\end{equation}
and would also expect $m_{\nu_{\tau}} \sim \left | C \right |$,
although if $\left | \frac{2X_1X_2}{A} \right | \gapprox \left | C \right |$
then $X_1$ and $X_2$ would also contribute to $m_{\nu_{\tau}}$.
Since we expect $m_{\nu_{\tau}}$ to be heavily suppressed
we would also expect:
\begin{equation}
\Delta m_{12} \sim
\max\{\left |B_1 \right |, \left |B_2 \right |\} >
\left | C \right |.
\end{equation}

We also have a constraint on the amplitude of double beta decay
expected in models of this type. This amplitude is proportional
to:
\begin{equation}
\langle m \rangle \sim (M_{\nu})_{11} = B_1
\end{equation}
It follows that $\left |\langle m \rangle \right | \lapprox \Delta m_{12}$
 and we may use
$\Delta m_{12} \lapprox 0.1 m_1$ together with
$\Delta m^2_{12} \sim 10^{-2} \ \mathrm{eV}^2$
to obtain:
\begin{equation}
\left | \langle m \rangle \right | \lapprox 0.02 \ \mbox{eV}
\end{equation}
which compares with the experimental bound \cite{betadecay, beta} of:
\begin{equation}
\left | \langle m \rangle \right | < (0.6 - 1.6) \ \mbox{eV}.
\end{equation}
Planned experiments \cite{betadecay, NEMO}
are expected to reach a sensitivity of
$\left | \langle m \rangle \right | \sim (0.1 - 0.3)  \ \mathrm{eV}$.

\section{An Explicit Model}
\label{model}

The AGUT model provides an example of one model of the type discussed
which can generate neutrino masses and mixings
of the form required by the previous section, yielding
a solution to the atmospheric and solar neutrino problems.
This model has an extended gauge group
\begin{equation}
G = SMG_1\otimes SMG_2 \otimes SMG_3 \otimes U(1)_f
\end{equation}
where $SMG_i = SU_i(3) \otimes SU_i(2) \otimes U_i(1)$.
This group breaks down at the Planck scale ($M_{Planck} \sim 10^{19}
\ \mathrm{GeV}$)
to the diagonal subgroup SMG of $SMG^3$, identified as the usual
SM gauge group, with the
$U(1)_f$ being totally broken. The fermions of the i'th
generation are put into the same representations under
$SMG_i$ as their usual SM representation, and are
trivial under the other two $SMG_j$s. Their charges under
$U(1)_f$ are then determined by anomaly cancellation requirements.

Four Higgs fields $S, W, T$ and $\xi$ (in addition to
the Weinberg-Salam Higgs field, $\phi_{WS}$) and their
representations under G were chosen in \cite{SMG3U1, fgns}
to break G down to the usual SM group and generate a realistic
charged fermion spectrum. The Higgs field $S$ was chosen to
have a VEV $\langle S \rangle = 1$, in units of $M_{Planck}$,
and the other VEVs were determined by a fit to the quark-lepton
masses and quark mixing angles:
\begin{equation}
\langle W\rangle  =  0.179, \quad   \label{Wvev}
\langle T\rangle  =  0.071, \quad   \label{Tvev}
\langle \xi\rangle  =  0.099 \label{xivev}
\end{equation}
This spectrum leads to the following charged lepton mass matrix,
where we ignore CP violating phases:
\begin{eqnarray}
M_E  & \sim & \langle \phi_{WS} \rangle \left ( \begin{array}{ccc}
        \langle W \rangle \langle T \rangle^2 \langle \xi \rangle^2 &
        \langle W \rangle T \rangle^2 \langle \xi \rangle^3 &
                \langle W \rangle \langle T \rangle^4 \langle\xi \rangle \\
        \langle W \rangle \langle T \rangle^2 \langle \xi \rangle^5 &
        \langle W \rangle \langle T \rangle^2 &  \langle W \rangle \langle
         T \rangle^4 \langle \xi \rangle^2 \\
        \langle W \rangle \langle T \rangle^5 \langle \xi \rangle^3 &
        \langle W \rangle^2 \langle T \rangle^4 & \langle W \rangle \langle T
\rangle
                        \end{array} \right )
\label{H_E}
\end{eqnarray}

However, as we noted in \cite{fgns}, in order to generate
 neutrino masses of the right
size it is necessary to introduce a new scale, since
the see-saw scale $\frac{\langle \phi_{WS} \rangle^2}{M_{Planck}}
\sim 3 \times 10^{-6} \ \mathrm{eV}$ is too small.
Here we do this by introducing a triplet (under $SU(2)$ in
the SM) Higgs field $\Delta$ to
generate the neutrino masses and
choose the Abelian charges of $\Delta$  so that
$\left(M_{\nu}\right)_{12} = \left(M_{\nu}\right)_{21}$ is unsuppressed
giving:
\begin{equation}
\left(\frac{y_1}{2}, \frac{y_2}{2}, \frac{y_3}{2}, y_f \right)
= \left(-\frac{1}{2}, -\frac{1}{2}, 0, 0 \right)
\end{equation}
where $\frac{y_i}{2}$ and $y_f$ are the charges under $U(1)_i$ and $U(1)_f$.
This Higgs field $\Delta$ is a doublet under $SU(2)_1$ and $SU(2)_2$,
but a singlet under all the other non-Abelian groups.
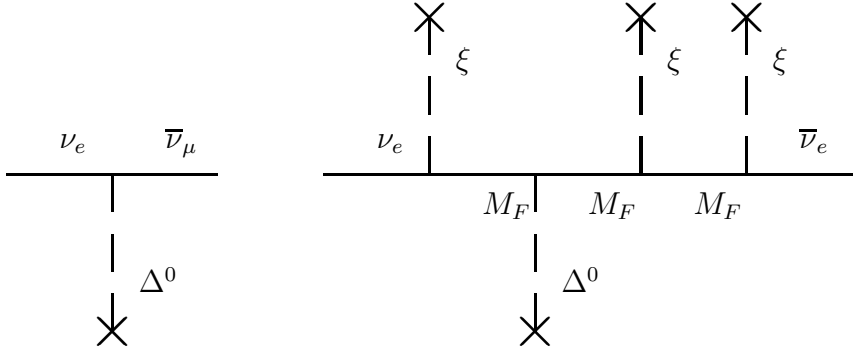
\begin{figure}
\begin{picture}(39000,13000)
\THICKLINES

\drawline\fermion[\E\REG](3000,7000)[4000]
\global\advance \pmidy by 1000
\put(\pmidx,\pmidy){${\nu}_e$}

\drawline\fermion[\E\REG](7000,7000)[4000]
\global\advance \pmidy by 1000
\put(\pmidx,\pmidy){$\overline{\nu}_{\mu}$}

\drawline\scalar[\S\REG](7000,7000)[3]
\global\advance \pmidx by 1000
\global\advance \pmidy by -1500
\put(\pmidx,\pmidy){$\Delta^0$}
\global\advance \scalarbackx by -530
\global\advance \scalarbacky by -530
\drawline\fermion[\NE\REG](\scalarbackx,\scalarbacky)[1500]
\global\advance \scalarbacky by 1060
\drawline\fermion[\SE\REG](\scalarbackx,\scalarbacky)[1500]

\drawline\fermion[\E\REG](15000,7000)[4000]
\global\advance \pmidy by 1000
\put(\pmidx,\pmidy){${\nu}_e$}

\drawline\fermion[\E\REG](19000,7000)[4000]
\global\advance \pmidy by -1500
\put(\pmidx,\pmidy){$M_F$}

\drawline\fermion[\E\REG](23000,7000)[4000]
\global\advance \pmidy by -1500
\put(\pmidx,\pmidy){$M_F$}

\drawline\fermion[\E\REG](27000,7000)[4000]
\global\advance \pmidy by -1500
\put(\pmidx,\pmidy){$M_F$}

\drawline\fermion[\E\REG](31000,7000)[4000]
\global\advance \pmidy by 1000
\put(\pmidx,\pmidy){$\overline{\nu}_e$}

\drawline\scalar[\S\REG](23000,7000)[3]
\global\advance \pmidx by 1000
\global\advance \pmidy by -1500
\put(\pmidx,\pmidy){$\Delta^0$}
\global\advance \scalarbackx by -530
\global\advance \scalarbacky by -530
\drawline\fermion[\NE\REG](\scalarbackx,\scalarbacky)[1500]
\global\advance \scalarbacky by 1060
\drawline\fermion[\SE\REG](\scalarbackx,\scalarbacky)[1500]

\drawline\scalar[\N\REG](19000,7000)[3]
\global\advance \pmidx by 1000
\global\advance \pmidy by 1000
\put(\pmidx,\pmidy){$\xi$}
\global\advance \scalarbackx by -530
\global\advance \scalarbacky by -530
\drawline\fermion[\NE\REG](\scalarbackx,\scalarbacky)[1500]
\global\advance \scalarbacky by 1060
\drawline\fermion[\SE\REG](\scalarbackx,\scalarbacky)[1500]

\drawline\scalar[\N\REG](27000,7000)[3]
\global\advance \pmidx by 1000
\global\advance \pmidy by 1000
\put(\pmidx,\pmidy){$\xi$}
\global\advance \scalarbackx by -530
\global\advance \scalarbacky by -530
\drawline\fermion[\NE\REG](\scalarbackx,\scalarbacky)[1500]
\global\advance \scalarbacky by 1060
\drawline\fermion[\SE\REG](\scalarbackx,\scalarbacky)[1500]

\drawline\scalar[\N\REG](31000,7000)[3]
\global\advance \pmidx by 1000
\global\advance \pmidy by 1000
\put(\pmidx,\pmidy){$\xi$}
\global\advance \scalarbackx by -530
\global\advance \scalarbacky by -530
\drawline\fermion[\NE\REG](\scalarbackx,\scalarbacky)[1500]
\global\advance \scalarbacky by 1060
\drawline\fermion[\SE\REG](\scalarbackx,\scalarbacky)[1500]

\end{picture}
\vskip .3cm
\caption{Example Feynman diagrams for neutrino mass.
The crosses indicate the couplings of the Higgs fields to the vacuum,
and $M_F \simeq M_{Planck}$}
\label{nutriplet}
\end{figure}
It generates the neutrino mass matrix:
\begin{eqnarray}
M_{\nu}  \sim  \langle {\Delta}^0\rangle\left ( \begin{array}{ccc}
        \langle \xi \rangle^3 &  1 &
        \langle T \rangle^3 \langle \xi \rangle^2 \\
        1 & \langle \xi \rangle^3 & \langle T \rangle^3 \langle \xi \rangle \\
        \langle T \rangle^3 \langle\xi \rangle^2 &
        \langle T \rangle^3 \langle \xi \rangle &
        \langle T \rangle^3 \langle W \rangle^3 \langle \xi \rangle
                        \end{array} \right )
\end{eqnarray}
by Feynman diagrams such as those in fig.~\ref{nutriplet}.
This matrix is clearly of the form required from the previous section,
with $\left (M_{\nu} \right )_{12}$ dominating the matrix to give
nearly degenerate electron and muon neutrino masses.
The diagonal element $\left (M_{\nu} \right )_{33}$ corresponds to
$m_{\nu_{\tau}}$ and is heavily suppressed, so the tau neutrino is
very light in this model and, in fact, the neutrino masses span 7
orders of magnitude.

Diagonalisation of the lepton mass matrices $M_E$ and $M_{\nu}$
then gives a lepton
mixing matrix:
\begin{eqnarray}
U  \sim \left ( \begin{array}{ccc}
\frac{1}{\sqrt{2}} & -\frac{1}{\sqrt{2}} &
\frac{\langle T\rangle^3\langle \xi\rangle}{\sqrt{2}} \\
\frac{1}{\sqrt{2}} & \frac{1}{\sqrt{2}} &
\frac{\langle T\rangle^3\langle \xi\rangle}{\sqrt{2}} \\
-\langle T\rangle^3\langle \xi\rangle &
\langle T\rangle^3\langle \xi\rangle^2 & 1
                        \end{array} \right )
\sim \left ( \begin{array}{ccc}
        0.71 & -0.71 & 3 \times 10^{-5} \\
        0.71 & 0.71 & 3 \times 10^{-5} \\
        -4 \times 10^{-5} & 4 \times 10^{-6} & 1
                        \end{array} \right ) \label{U}
\end{eqnarray}
As we can see from this mixing matrix the tau neutrino is virtually
unmixed in this model, with the mixing being much less than
the constraints given by CDHS \cite{CDHS} ($\sin^2 2\theta_{\mu\tau} \lapprox
0.1$),
 and the sensitivities
of CHORUS \cite{CHORUS} and NOMAD \cite{NOMAD},
which for $\Delta m^2_{\nu_{\mu}\nu_{\tau}} \sim
15 \ \mathrm{eV}^2$ are sensitive to
$\sin^2 2\theta_{\mu\tau} > 10^{-3}$. So we
 essentially have two neutrino mixing
between the electron and muon neutrinos with
$\sin^2(2\theta) \sim 1$.

We choose the mass scale $\langle \Delta^0 \rangle \approx 2 \ \mathrm{eV}$
in order to get an appropriate value for $\Delta m^2_{12}$ giving:
\begin{eqnarray}
\Delta m_{12}^2 \sim 2 {\langle {\Delta}^0\rangle}^2
{\langle \xi \rangle}^3
\approx 8 \times 10^{-3} \ \mbox{eV}^2\\
m_{1} \approx m_{2}
\approx 2 \ \mbox{eV},
\ \ m_{\nu_{\tau}} \approx  2 \times 10^{-7} \ \mbox{eV}
\end{eqnarray}
which, since we have almost maximal mixing between ${\nu}_e$ and
${\nu}_{\mu}$, is suitable for the
solution to both the solar and atmospheric neutrino problems.
As expected the model is incompatible with the LSND result,
and it also gives masses suitable for hot dark matter, with
$\sum m_\nu \sim 4 \ \mathrm{eV}$.
The model also makes a prediction for neutrinoless double
beta decay  of $\left | \langle m \rangle \right |
 \sim \langle \Delta^0 \rangle
\xi^3 \sim 2 \times 10^{-3} \ \mathrm{eV}$,
which as expected is much less than values accessible by current
or planned experiments.

\section{Conclusions}
\label{conclusions}

We have found that, in models which give a natural solution
to the fermion mass hierarchy problem, the only way of naturally explaining
both
the solar and atmospheric neutrino problems is if both
effects are due to nearly maximal
$\nu_e - \nu_{\mu}$ mixing with $\Delta m^2_{12} \sim 10^{-2} \
\mathrm{eV}^2$; this leads to our prediction of an electron neutrino flux
suppression factor of $\frac{1}{2}$ in all solar neutrino experiments.
 The electron and muon neutrinos are nearly degenerate
with masses of order $1 \ \mathrm{eV}$, and are therefore likely to
be hot dark matter candidates, whilst the tau neutrino
is much lighter and only slightly mixed.

The prospects for examining this scenario experimentally in the near future
are very good; reactor experiments on $\overline{\nu}_e$
survival rates, such as CHOOZ \cite{CHOOZ} and PALO-VERDE \cite {Palo Verde},
will be
able to reach $\Delta m^2_{12} \sim 10^{-3} \ \mathrm{eV}^2$, and we
would expect to see a strong signal of neutrino
oscillations there.
The LSND result will also prove to be unfounded
if our scenario is correct. It should be noted that a characteristic of
this scenario is that we would not expect to see either seasonal or
day/night effects from the solar neutrinos.
It is harder to verify the tau neutrino mixing since it is much weaker; however
if the muon neutrino is heavier than a few eV then in some models the tau
neutrino
mixing may be sufficiently large (e.g. the mixing could be of
order $\frac{m_{\mu}}{m_{\tau}}$ coming from $M_E$)
to give a signal at the CERN experiments CHORUS and NOMAD.
We conclude that we must have essentially unique neutrino masses and
mixing, with large $\nu_e - \nu_{\mu}$ mixing, which can be confirmed
or excluded by experiment within the next few years.

{}From the theoretical point of view the above neutrino mass
scale implies some new physics between the electroweak scale and the
Planck scale. So the assumption of a total ``desert'' up to about
one order of magnitude below $M_{Planck}$, as in the AGUT model,
is not consistent with our interpretation of neutrino phenomenology.
Either some intermediate mass see-saw fermions or a Higgs field like
$\Delta$ is required. The latter could acquire a vacuum expectation
value $\langle \Delta^0 \rangle \sim 1$ eV, via its interaction
with two Weinberg-Salam Higgs fields $\phi_{WS}$ and the other Higgs fields
$W$, $T$, $\xi$ and $S$; but only if, for some as yet unknown reason,
it has a very small coefficient of $\Delta^2$ in the Higgs
potential compared to $M_{Planck}^2$.

\section*{Acknowledgements}

HBN acknowledges funding from INTAS 93-3316, EF contract SCI 0340 (TSTS)
and Cernf{\o}lgeforskning. CF acknowledges funding from INTAS 93-3316
and PPARC GR/J21231. MG is grateful for a PPARC studentship.

\end{document}